\documentclass[11pt]{article}
\usepackage{graphicx}

\newcommand{\BABARPubYear}    {04}

\newcommand{\BABARConfNumber} {24}
\newcommand{\SLACPubNumber} {10584}

\input pubboard/babarsym

\long\def\inst#1{\par\nobreak\kern 4pt\nobreak
    {\it #1}\par\vskip 10pt plus 3pt minus 3pt}

\begin{document}
{\pagestyle{empty}

\begin{flushright}
\babar-CONF-\BABARPubYear/\BABARConfNumber \\
SLAC-PUB-\SLACPubNumber \\
\end{flushright}

\par\vskip 5cm

\begin{center}
\Large \bf Measurement of exclusive \boldmath \B \unboldmath decays to charmonium and 
\boldmath \kaon \unboldmath or 
\boldmath \Kstar \unboldmath branching fractions with the \babar\ detector.
\end{center}
\bigskip

\begin{center}
\large The \babar\ Collaboration\\
\mbox{ }\\
\today
\end{center}
\bigskip \bigskip

\begin{center}
\large \bf Abstract
\end{center}
We report preliminary results on the measurement of branching fractions of exclusive decays of 
neutral and 
charged \B mesons into two-body final states containing a charmonium state and a light strange 
meson. The charmonium mesons considered are \jpsi, \psitwos and \chicone, and the light 
mesons are either \kaon or \Kstar. We use a sample of about 124 million \BB events 
collected with the \babar\ detector at the PEP-II storage ring at the Stanford Linear 
Accelerator Center.

\vfill
\begin{center}

Submitted to the 32$^{\rm nd}$ International Conference on High-Energy Physics, ICHEP 04,\\
16 August---22 August 2004, Beijing, China

\end{center}

\vspace{1.0cm}
\begin{center}
{\em Stanford Linear Accelerator Center, Stanford University, 
Stanford, CA 94309} \\ \vspace{0.1cm}\hrule\vspace{0.1cm}
Work supported in part by Department of Energy contract DE-AC03-76SF00515.
\end{center}

\newpage
} 

\begin{center}
\small

The \babar\ Collaboration,
\bigskip

%
B.~Aubert,
R.~Barate,
D.~Boutigny,
F.~Couderc,
J.-M.~Gaillard,
A.~Hicheur,
Y.~Karyotakis,
J.~P.~Lees,
V.~Tisserand,
A.~Zghiche
\inst{Laboratoire de Physique des Particules, F-74941 Annecy-le-Vieux, France }
% ...

\end{center}\newpage

\section{Introduction}
\label{sec:Introduction}
\begin{figure}[!htb]
\begin{center}
\includegraphics{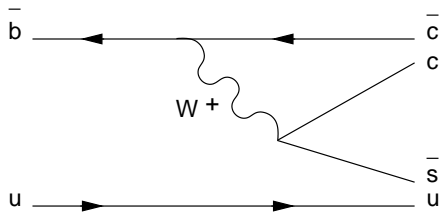}
\caption{Tree level diagram of a \B meson decaying into a charmonium 
state and a kaon.}
\label{fig:decay}
\end{center}
\end{figure}

Fully hadronic decays of \B mesons have proven to be an effective laboratory to 
study and provide tests of the theory of heavy quarks as well as the dynamics 
of strong interactions in heavy meson systems. The tree level diagram of the 
decays under study is shown in Figure \ref{fig:decay}. 

The dynamics of the decay is expected to be highly affected by strong interactions 
effects, especially by the long distance non-pertubative aspect of QCD. There are 
various phenomenological approaches to treat these decays, which provide different 
estimates for the branching fractions (see \cite{ref:oldanalysis} and references 
[2-12] therein). 

Charge asymmetry measurements can be a powerful tool for seeking new 
physics. The Standard Model predicts small direct \CP violation \cite{ref:smallasymm}, 
thus large charge asymmetries would indicate new physics \cite{ref:newphys}.

The list of the branching fractions measured and decay modes considered in this 
paper is shown in Table \ref{tab:list}.

\begin{table}[h]
\caption{
Branching fractions and decay modes considered in this analysis.}
\begin{center}
\begin{tabular}{l|ll} \hline \hline
  Decay Channel & Secondary decay mode & \\ \hline
  \Bz \ra \jpsi \Kstarz & \Kstarz \ra \Kp \pim , \KS \piz &  \jpsi \ra \ellell\\
  \Bp \ra \jpsi \Kstarp & \Kstarp \ra \Kp \piz , \KS \pip &  \KS \ra \pipi  \\
  \Bz \ra \jpsi \KS &   & \piz \ra \gaga \\
  \Bp \ra \jpsi \Kp &   & \\ \hline
  \Bz \ra \psitwos \Kstarz & \Kstarz \ra \Kp \pim , \KS \piz & \psitwos \ra \ellell \\
  \Bp \ra \psitwos \Kstarp & \Kstarp \ra \Kp \piz , \KS \pip & \KS \ra \pipi  \\
  \Bz \ra \psitwos \KS &  & \piz \ra \gaga \\
  \Bp \ra \psitwos \Kp &  & \\ \hline
  \Bz \ra \chicone \Kstarz & \Kstarz \ra \Kp \pim , \KS \piz  & \chicone \ra \jpsi \g \\
  \Bp \ra \chicone \Kstarp & \Kstarp \ra \Kp \piz , \KS \pip  & \jpsi \ra \ellell \\
  \Bz \ra \chicone \KS & & \KS \ra \pipi  \\
  \Bp \ra \chicone \Kp & & \piz \ra \gaga \\ \hline \hline
\end{tabular}
\end{center}
\label{tab:list}
\end{table}

\section{The \babar\ detector and dataset}
\label{sec:babar}

The data used in this analysis were collected with the \babar\ detector at the 
\pep2\  asymmetric \epem storage ring from 1999 to 2003. This represents a 
total integrated luminosity of 112.4 fb$^{-1}$ taken on the $\Upsilon$(4S) resonance, 
producing a sample of 123.95 million \BB events.

The \babar\ detector is described elsewhere~\cite{ref:babar}. Surrounding the 
interaction point, a 5 layer double-sided silicon vertex tracker (SVT) provides 
precise reconstruction of track angles and \B decay vertices. A 40 layer drift 
chamber (DCH) provides measurements of the transverse momenta of charged particles. 
An internally reflecting ring-imaging Cherenkov detector (DIRC) is used for 
particle identification. 
A CsI(Tl) crystal electromagnetic calorimeter (EMC) is used to detect photons and electrons. 
The calorimeter is surrounded by a 1.5T magnetic field. The flux return is 
instrumented with resistive plate chambers (IFR) used for muon and neutral hadron identification.

\section{Analysis Method}
\label{sec:Analysis}
Multihadron events are selected by demanding a minimum of three
reconstructed charged tracks in the polar-angle range $0.41 <
\theta_{lab} < 2.54$\rad. Charged tracks must be reconstructed in the
DCH and are required to originate at the nominal beamspot, within
1.5\cm in the plane transverse to the beam and 10\cm along the beam.
Events are required to have a primary vertex 
within 0.5\cm of the average position of the interaction point in the plane transverse to
the beamline, and within 6\cm longitudinally.

Charged tracks used in this analysis are required to include at least 12 DCH hits, to 
have a transverse momentum \pt$>$100 \mevc. 

Photons are reconstructed from EMC clusters. The radial energy profile (LAT) \cite{ref:lat} 
of the cluster is used to discriminate electromagnetic from hadronic clusters. Photons are 
required to have a minimum energy of 30 MeV, a radial energy profile less than 0.8, 
and to be in the fiducial volume $0.41 < \theta < 2.41$ rad.

Electron candidates are selected using information from the EMC (radial energy profile 
and Zernike moment $A_{42}$ \cite{ref:zernike}) , the ratio of the energy measured in the 
EMC to the 
momentum measured by the tracking system (E/p), energy loss (dE/dx) in the drift chamber 
and the Cherenkov angle measured in the DIRC. Electrons are also required to be 
in the fiducial volume $0.41 < \theta < 2.41$ rad.

Muon candidates are selected using information from the EMC (energy deposition 
consistent with a minimum ionizing particle) and the distribution of hits in the 
IFR. Muons are required to be in the fiducial volume $0.3 < \theta < 2.7$ rad.

The charged kaon and pion candidates are selected using information from the 
energy loss in the SVT and DCH, and the Cherenkov angle measured in the DIRC. Kaon 
candidates are required to be in the fiducial volume $0.45 < \theta < 2.45$ rad.

\vskip 0.5cm
\noindent
The next step in the analysis is to combine tracks and/or neutral clusters 
to form candidates. If a particle decays through an intermediate state, this is 
constrained to its known mass, except for the \Kstar. The selection has been 
optimized by maximizing the ratio $S/ \sqrt{S+B}$, where S and B are respectively 
the number of expected signal and background events obtained from GEANT4-based Monte Carlo 
simulation after the selection.

The \jpsi candidates are required to have an invariant mass $2.95 < M_{\epem} < 3.14$
\gevcc and $3.06 < M_{\mumu} < 3.14$ for \jpsi \ra \epem and \jpsi \ra \mumu decays 
respectively. 

The \psitwos candidates are required to have an invariant mass 
$3.44 < M_{\epem} < 3.74$ \gevcc and $3.64 < M_{\mumu} < 3.74$ \gevcc for \psitwos \ra \epem 
and \psitwos \ra \mumu decays respectively. 

For \jpsi\to\epem and \psitwos\to\epem decays, electron candidates are combined with photon 
candidates in order to recover some of the energy lost through bremsstrahlung. Photons are 
required to be within 35\mrad in polar angle from the electron track, and to have 
an azimuthal angle intermediate between the initial track direction (estimated by 
subtracting 50\mrad opposite to the bend direction of the reconstructed track) and 
the centroid of the EMC cluster arising from the track.

In the \chicone reconstruction (\chicone \ra \jpsi \g), \jpsi candidates are selected as 
described above. The associated \g has to fulfill the following requirements: radial 
energy profile less than 0.8, Zernike moment $A_{42}$ less than 0.15 and energy greater 
than 0.15 \gev. Furthermore, \chicone candidates are required to satisfy 
$0.35 < M_{\ellell \g} - M_{\ellell} < 0.45$ \gevcc.

The \piz \ra \gaga candidates are required to satisfy $0.113 < M_{\gaga} < 0.153$ \gevcc. 
The radial energy profile of both photons are required to be less than 0.8. The energy 
of the soft photon has to be greater than 0.050 \gev and the energy of the hard 
photon has to be greater than 0.150 \gev. 

The \KS \ra \pip \pim  candidates are required to satisfy $0.489 < M_{\pipi} < 0.507$ \gevcc. 
The following selection is also required: the \KS vertex has be more than 1mm from the charmonium 
vertex, and the angle in the x-y plane between the \KS momentum and the line joining the 
charmonium and \KS vertices has to be smaller than 0.2 rad.

The \Kstarz and \Kstarp candidates are respectively required to satisfy 
$0.796 < M_{\kaon \pi} < 0.996$ \gevcc and $0.792 < M_{\kaon \pi} < 0.992$ \gevcc. In addition, 
for channels having a \piz in the 
final state, the cosine of the angle between the \kaon momentum defined in the 
\Kstar rest frame and the \Kstar momentum defined in the \B rest frame has to be 
smaller than 0.8 (this helps in removing background coming from events with soft pions). 

\begin{figure}[!htb]
\begin{center}
\includegraphics[width=0.3\textwidth]{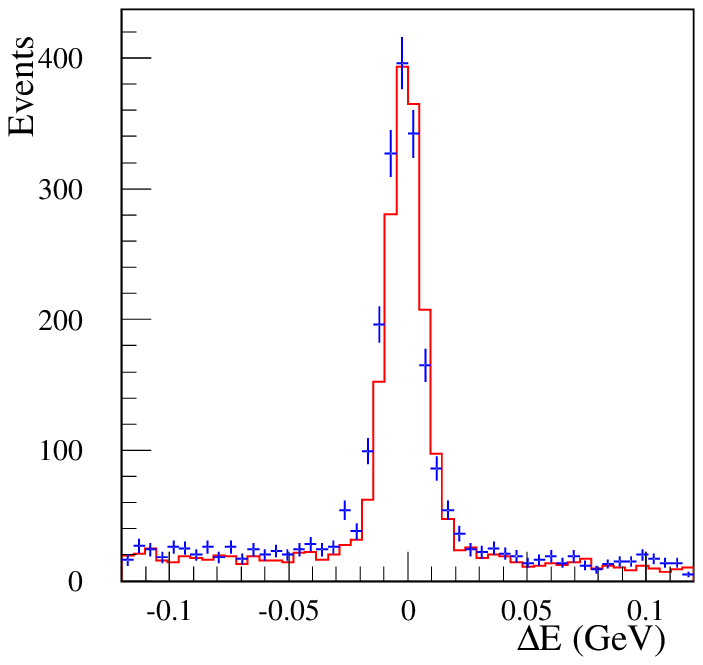}
\includegraphics[width=0.3\textwidth]{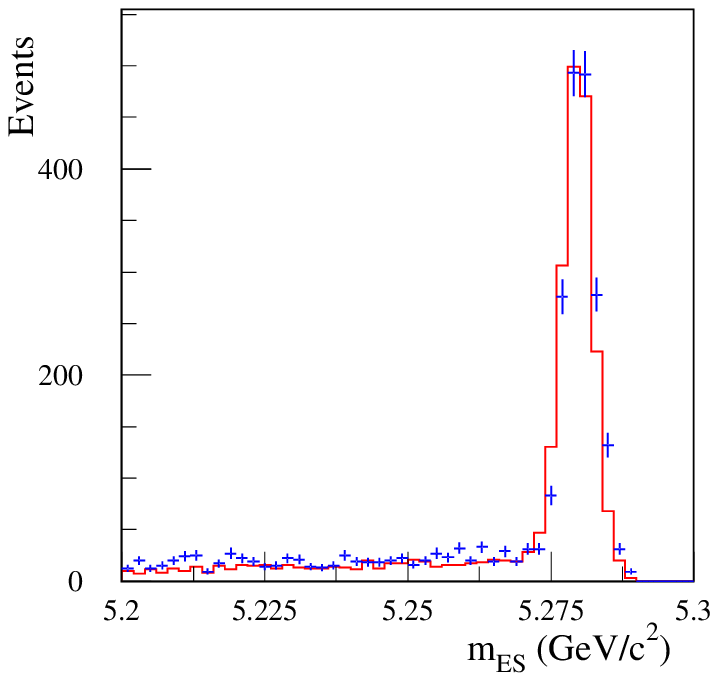}
\caption{\DeltaE and \mes distributions for the \B \ra \jpsi \Kstarz (\Kp \pim) 
channel. The blue points represent the Data and the histogram represents the Monte 
Carlo. An offset between the Monte Carlo and data distributions can be seen. It has 
been corrected.}
\label{fig:examplejpsikstar0kp}
\end{center}
\end{figure}

Finally, \B candidates are reconstructed by combining charmonium and kaon
meson candidates and are selected by the use of two kinematic variables: the difference 
between the reconstructed energy of the \B candidate and the beam energy in the 
center-of-mass frame $\DeltaE = E_B^*-E_{beam}^*$, and the beam energy 
substituted mass \mes, defined as $\mes \equiv \sqrt{E_{beam}^{*2}-{\bf p}_B^{*2}}$ 
(the $^*$ refers to quantities in the center of mass). For a true \B meson,
\DeltaE is expected to peak at zero, and the energy substituted mass \mes
should peak at the \B meson mass 5.279 \gevcc. Only one reconstructed \B meson is 
allowed per event. For events that have multiple candidates, the candidate having 
the smallest \DeltaE is chosen. Depending on the channel, around 10$\%$ of the 
candidates are removed by requesting a single \B meson per event. The analysis is performed 
in the \mes vs \DeltaE plane, defined as: $5.2 < \mes < 5.3$ \gevcc and 
$-0.12 < \DeltaE < 0.12$ \gev. As an example, Figure \ref{fig:examplejpsikstar0kp} shows 
the \DeltaE and \mes distributions for the \B \ra \jpsi \Kstarz (\Kp \pim) 
channel. 
We subsequently define a signal box region in the \mes vs \DeltaE plane, where 
the sensitivity is optimal. The signal box region is channel-dependent. For most of the 
channels, the signal regions are taken as the mean value $\pm 3 \sigma$ for both \DeltaE 
and \mes. For channels with less statistics (\psitwos \Kstar and \chicone \Kstar channels), 
the \mes signal region was taken as $5.27 < \mes < 5.29$ \gevcc, and the \DeltaE signal 
region was taken as $| \DeltaE | < 0.04$ \gev for channels with a \piz in the final 
state and $| \DeltaE | < 0.03$ \gev for the other channels.

\vskip 0.5cm
\noindent
The selection efficiencies for each mode are obtained from Monte Carlo and are 
given by the number of expected signal events divided by the total number of generated 
events. While the Monte Carlo has been tuned to be as realistic as possible, one still has 
to correct for residual differences between data and simulated events. We have therefore 
applied additional corrections to the selection efficiency coming from particle 
identification, neutral particle, tracking, and \KS corrections. 

\vskip 0.5cm
\noindent
The number of signal events $N_S$ is determined from the number of candidate events 
$N_{\rm cand}$ after subtracting the background. The \mes distribution within 
the \DeltaE signal region is fitted by an Argus function \cite{ref:argus} and a Gaussian, 
and both functions are subsequently integrated within the \mes signal region. 
The number of candidate events is given by the Gaussian integral. 
There are two components to the background: the 
combinatorial background and a peaking component (the component of the background that 
has \DeltaE and \mes distributions peaking at \DeltaE=0 \gev and \mes=5.279 \gevcc 
respectively). The combinatorial background is obtained by integrating the Argus function 
within the \mes and \DeltaE signal regions. The peaking 
component is obtained from Monte Carlo. There are two contributions to the peaking 
background. The first contribution is coming from feed-across
events which, in the case of the \jpsi \Kstarz (\KS \piz) reconstruction, for
instance, come from 
\jpsi \Kstarz (\Kp \pim), \jpsi \Kstarp (\KS \pip) and \jpsi \Kstarp (\Kp \piz). 
The second contribution is coming from inclusive charmonium. 
For each of the contributions, the \mes distribution is fitted 
within the \DeltaE signal region by an Argus and a Gaussian function, which are 
subsequently integrated within the \mes signal region. The amount of peaking 
background is given by the Gaussian integral.

\vskip 0.5cm
\noindent
The branching fractions are obtained from:

\begin{eqnarray}
  BF = {N_{S} \over N_{\BB} \times \epsilon \times f}
\end{eqnarray}

\noindent
where $N_{\BB}$ is the number of \BB events, $\epsilon$ is the selection 
efficiency and $f$ is the total secondary branching fraction. For channels with 
a \Kstar in the final state, the feed-across contribution, which depends on 
the branching fractions that are being measured, to the peaking background 
can be important. Therefore an iterative procedure has been employed in 
which the feed-across contribution is re-estimated at each iteration. The 
procedure converges quickly as the feed-across is a small fraction of the 
number of signal events. When allowed by the size of the data sample, 
the branching fractions have been measured for both \jpsi \ra \epem and 
\jpsi \ra \mumu decays separately.

\begin{figure}
  \center{
  \includegraphics[width=0.22\textwidth]{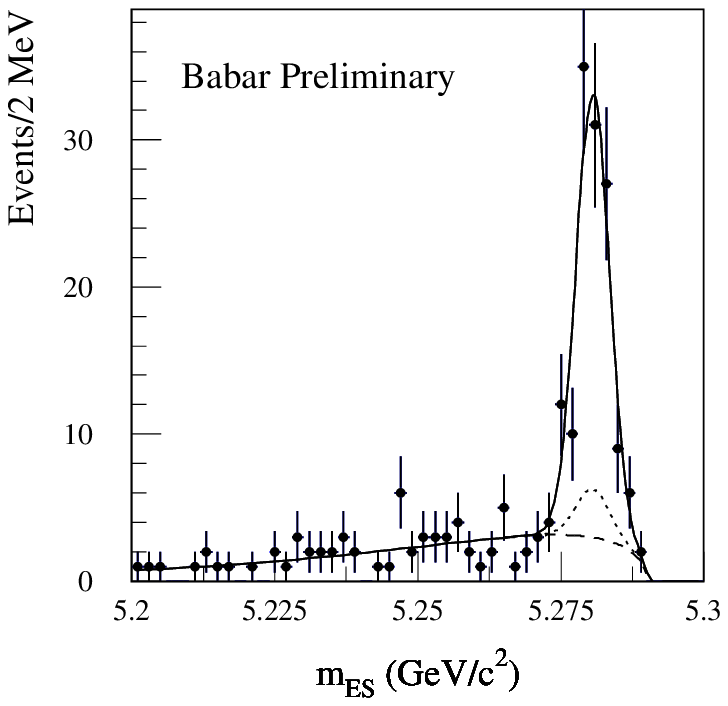}
  \includegraphics[width=0.22\textwidth]{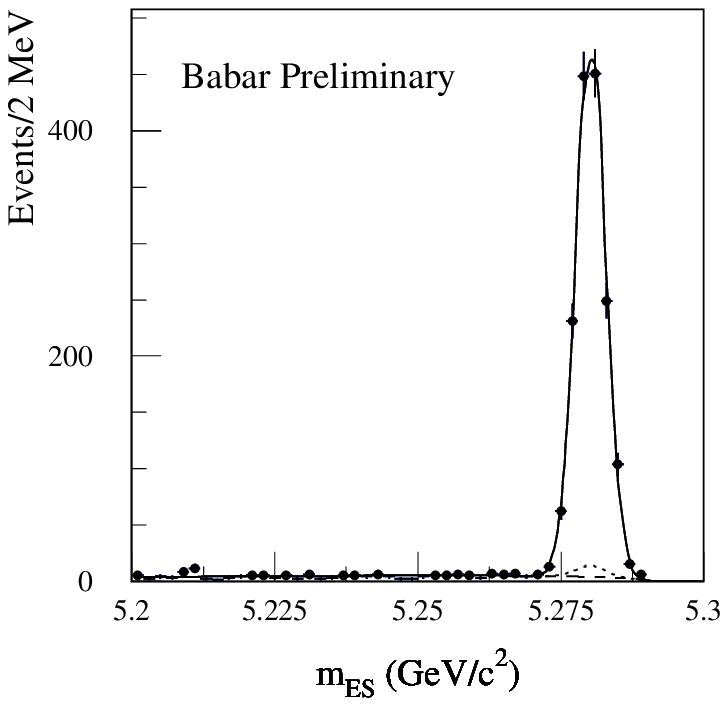}
  \includegraphics[width=0.22\textwidth]{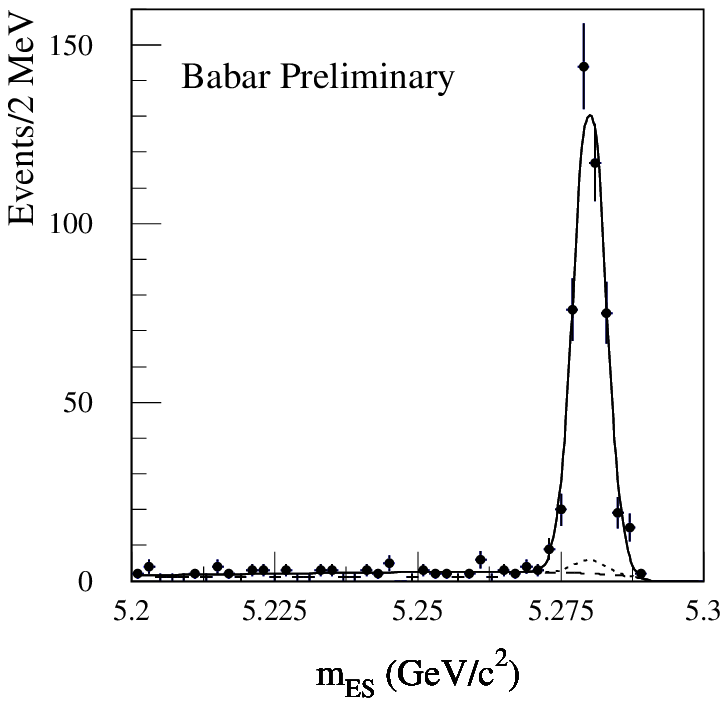}
  \includegraphics[width=0.22\textwidth]{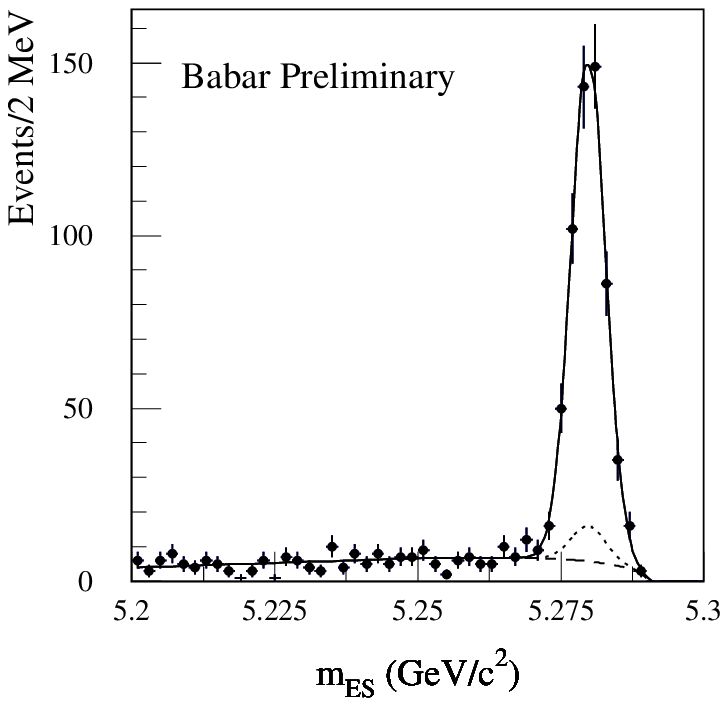}
  \includegraphics[width=0.22\textwidth]{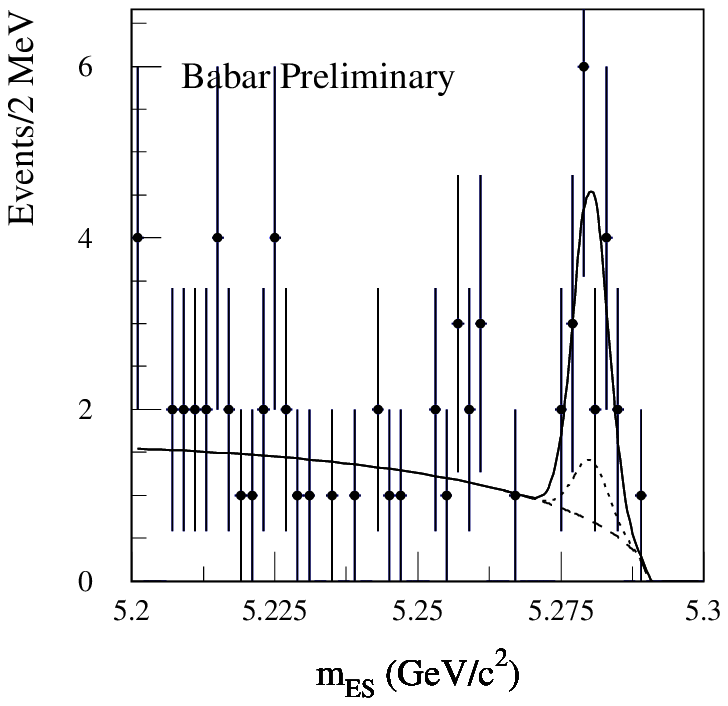}
  \includegraphics[width=0.22\textwidth]{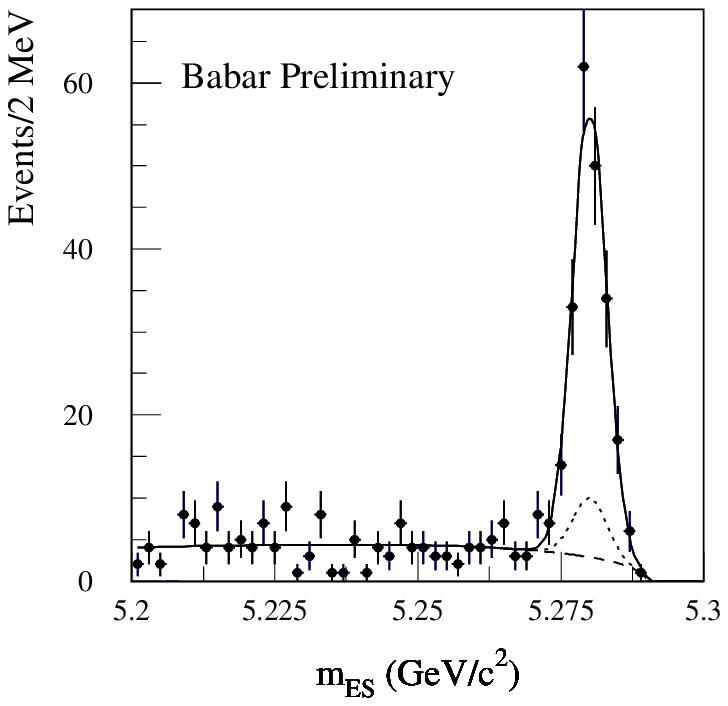}
  \includegraphics[width=0.22\textwidth]{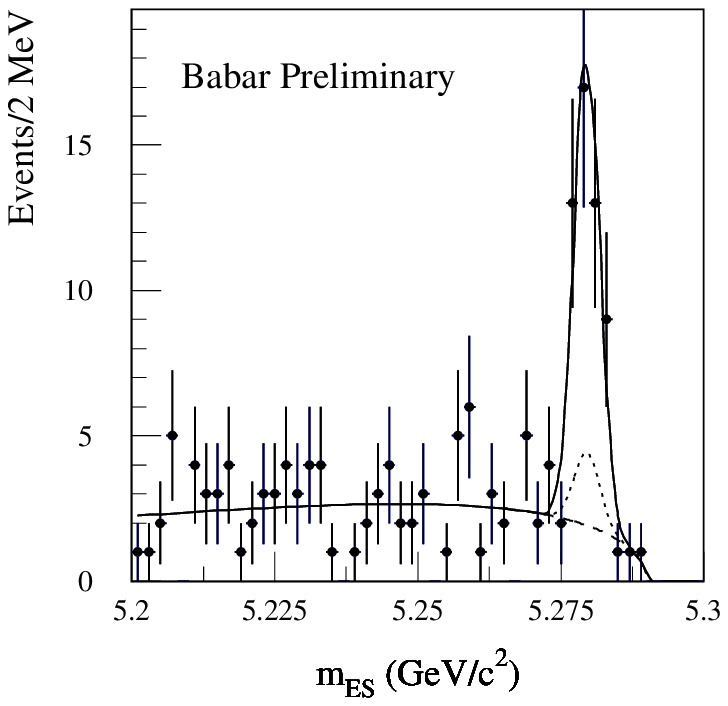}
  \includegraphics[width=0.22\textwidth]{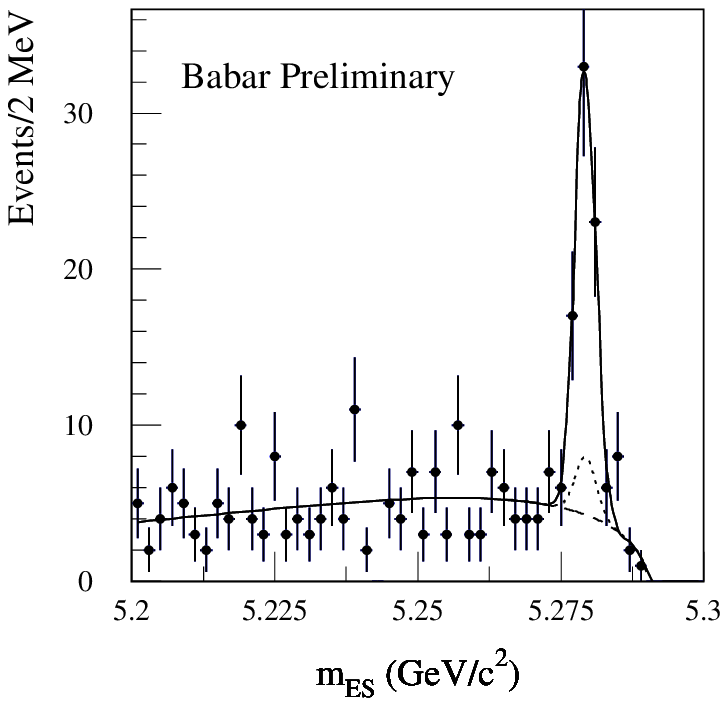}
  \includegraphics[width=0.22\textwidth]{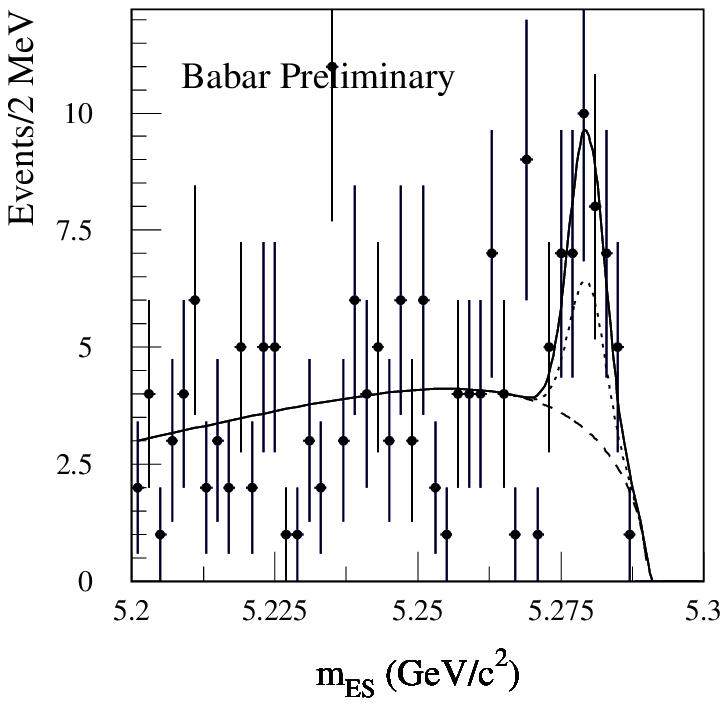}
  \includegraphics[width=0.22\textwidth]{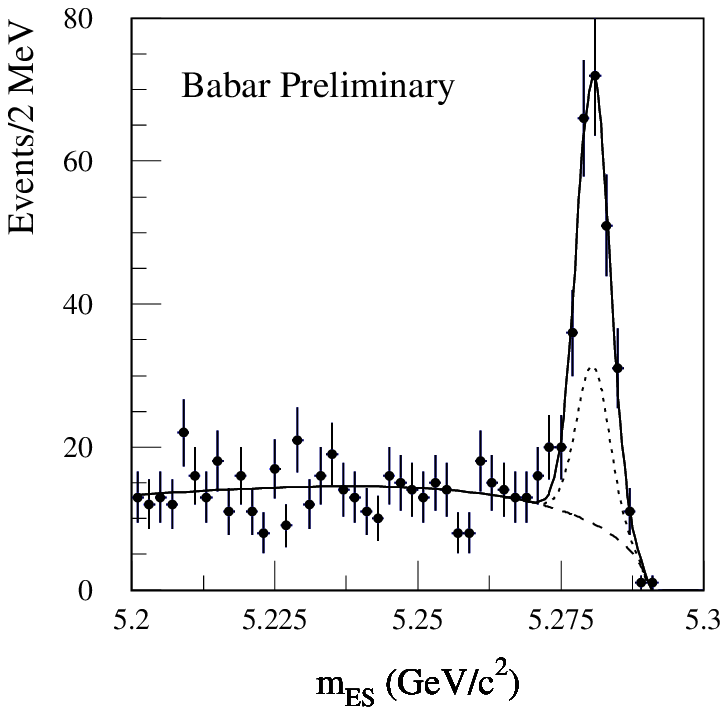}
  \includegraphics[width=0.22\textwidth]{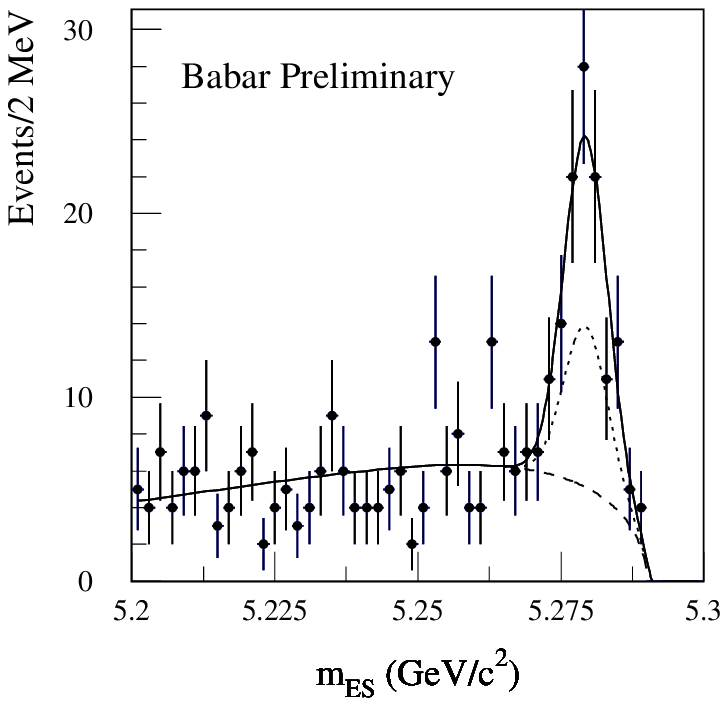}
  \includegraphics[width=0.22\textwidth]{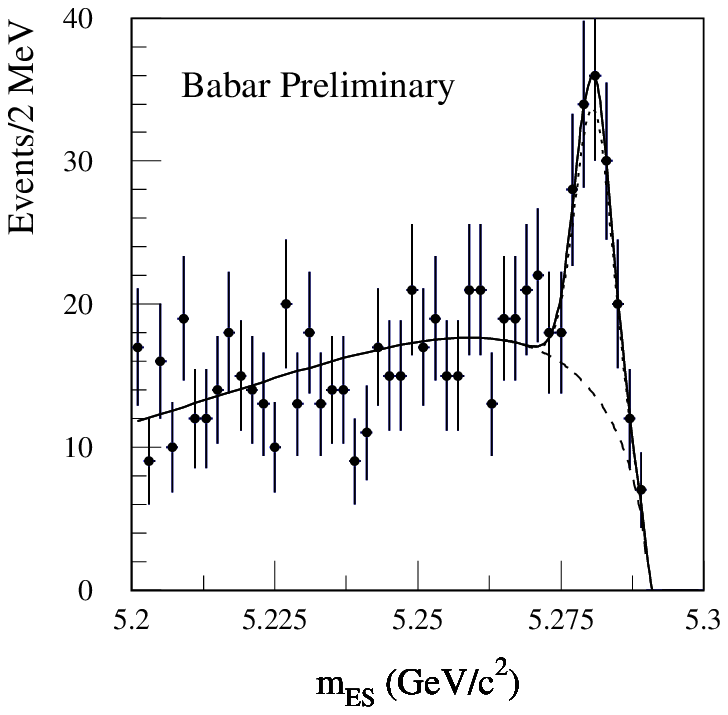}
 \caption{\label{fig:meskstar}
\mes distributions and fits within the \DeltaE\ {\tt Signal Box} region for 
\B \ra charmonium \Kstar channels. The top row represents the distributions for 
the \jpsi \Kstar channels, the middle row the \psitwos \Kstar channels, and the 
bottom row the \chicone \Kstar channel. From left to right, the columns show the 
distributions for the \Kstarz \ra \KS \piz, \Kstarz \ra \Kp \pim, \Kstarp \ra \KS \pip, 
and \Kstarp \ra \Kp \piz decay modes. The dashed lines show the combinatorial 
contribution to the background. The dotted lines show the peaking background 
contribution.}}
\end{figure}

\begin{figure}[!htb]
  \center{
  \includegraphics[width=0.3\textwidth]{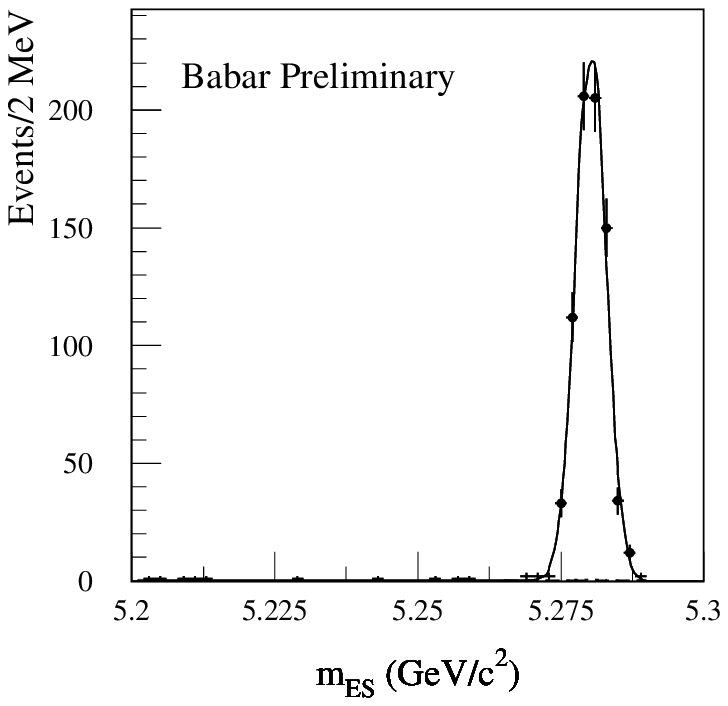}
  \includegraphics[width=0.3\textwidth]{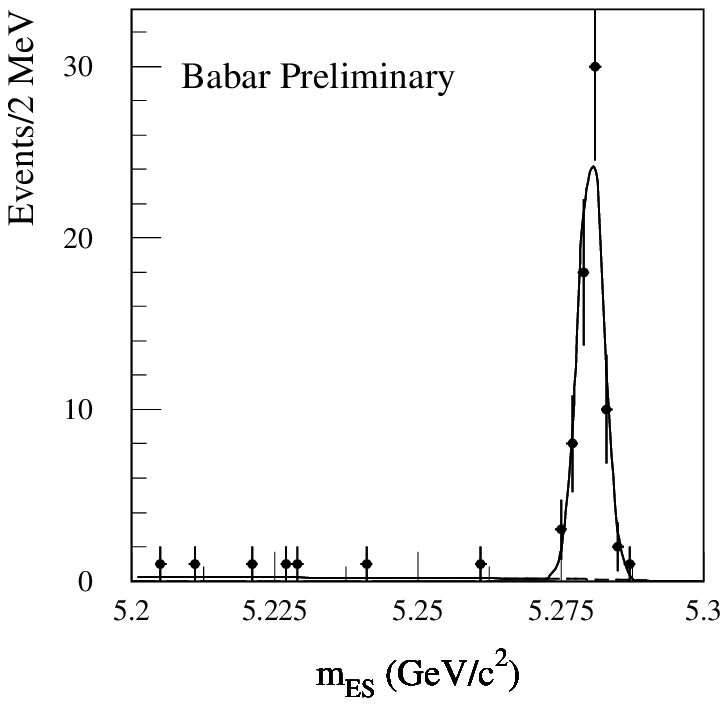}
  \includegraphics[width=0.3\textwidth]{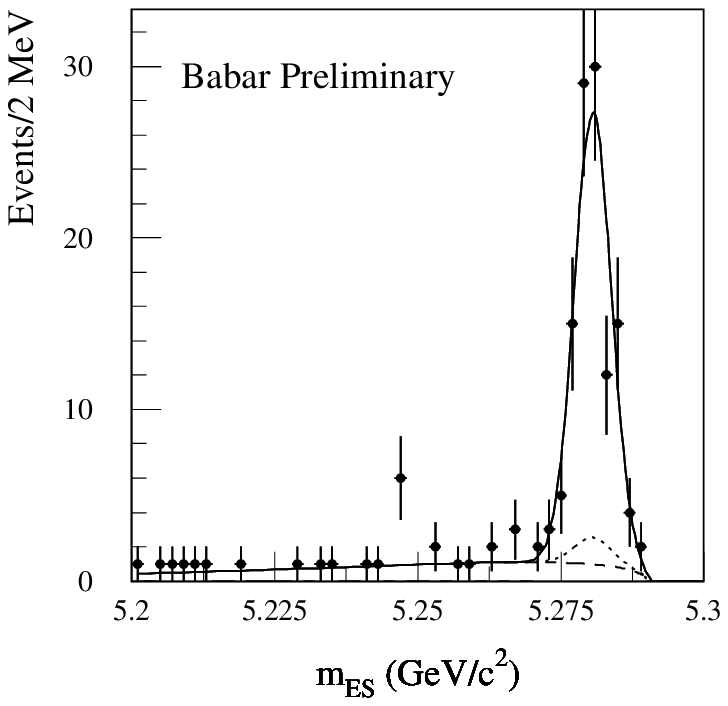}
  \includegraphics[width=0.3\textwidth]{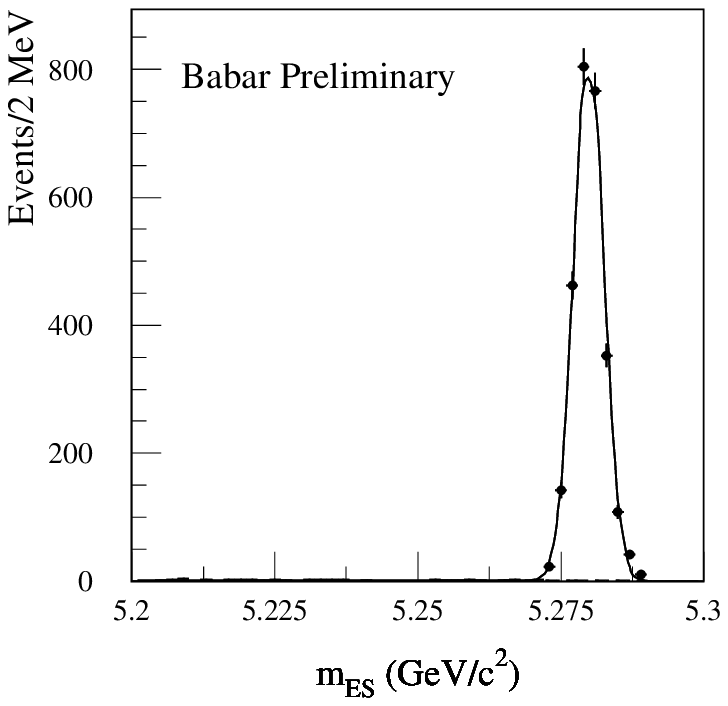}
  \includegraphics[width=0.3\textwidth]{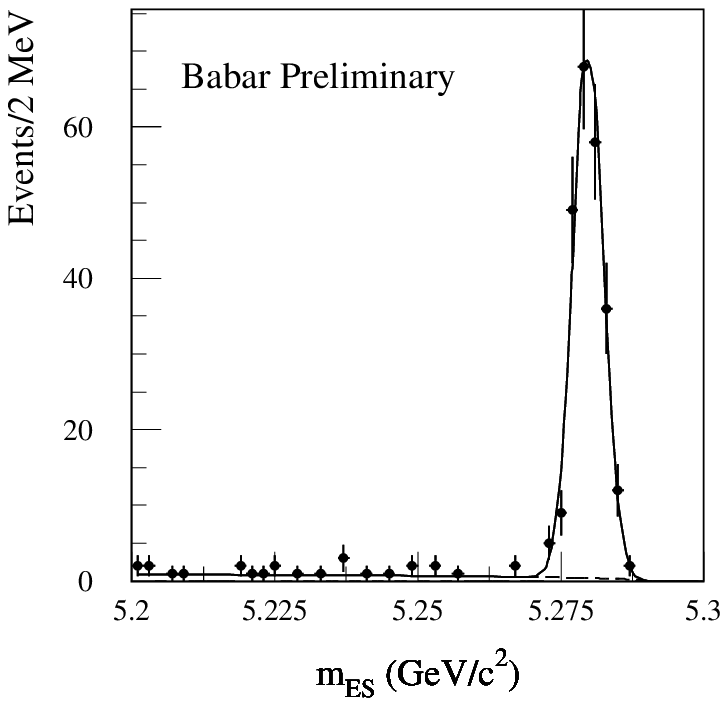}
  \includegraphics[width=0.3\textwidth]{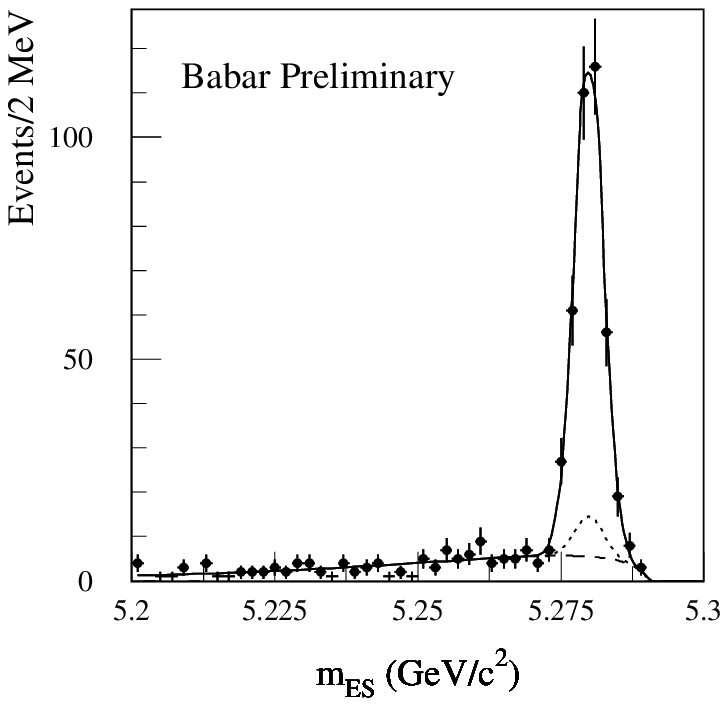}
 \caption{\label{fig:meskpks} 
\mes distributions and fits within the \DeltaE\ {\tt Signal Box} region for 
\B \ra charmonium \KS channels (top row) and \B \ra charmonium \Kp channels (bottom). 
From left to right, the columns show the distributions for the \jpsi,\psitwos and 
\chicone channels.}}
\end{figure}

The \mes distributions  within the \DeltaE\ signal region for candidate events 
are shown on Figures \ref{fig:meskstar} and \ref{fig:meskpks}.

\section{Systematic studies}
\label{sec:Systematics}

The systematic errors arise from the uncertainty on the number of \BB events, 
the secondary branching fraction, the estimate of the selection efficiency, and 
the knowledge of the background.

\begin{table}[!htb]
\caption{\label{tab:syste-jpsikstar}
Breakdown of contributions to the systematic errors for the \jpsi \Kstar channels. 
All values are expressed relative to the measured branching fractions, in percent.}
\begin{center}
\begin{tabular}{l|cc|cc|cc|cc} 
  & \multicolumn{2}{c}{\KS \piz}  & \multicolumn{2}{c}{\Kp \pim} & 
    \multicolumn{2}{c}{\KS \pim} & \multicolumn{2}{c}{\Kp \piz} \\

 &\epem&\mumu&\epem&\mumu&\epem&\mumu&\epem&\mumu \\ \hline

B counting & \multicolumn{2}{|c|}{1.1} & \multicolumn{2}{|c|}{1.1} & 
\multicolumn{2}{|c|}{1.1} & \multicolumn{2}{|c}{1.1} \\
Tracking & \multicolumn{2}{|c|}{2.6} & \multicolumn{2}{|c|}{5.2} & 
\multicolumn{2}{|c|}{3.9} & \multicolumn{2}{|c}{3.9} \\
Polarizarion & \multicolumn{2}{|c|}{7.37} & \multicolumn{2}{|c|}{3.85} & 
\multicolumn{2}{|c|}{4.85} & \multicolumn{2}{|c}{6.86} \\

\KS & \multicolumn{2}{|c|}{0.7} & \multicolumn{2}{|c|}{-} & 
\multicolumn{2}{|c|}{0.8} & \multicolumn{2}{|c}{-} \\

Neutral       & 5.72 & 5.27 & \multicolumn{2}{|c|}{-} & 
               \multicolumn{2}{|c|}{-} & 5.14 & 5.61 \\

Second BF     & 1.69 & 1.71 & 1.69 & 1.70 & 1.70 & 1.71 & 1.69 & 1.70 \\

PID           & 0.95 & 2.14 & 0.88 & 3.57 & 0.90 & 2.93 & 0.61 & 3.73 \\
Background    & 3.57 & 3.25 & 1.18 & 0.91 & 1.45 & 0.96 & 2.34 & 2.04 \\
MC statistics & 1.24 & 1.23 & 1.37 & 1.32 & 1.51 & 1.45 & 1.80 & 1.71 \\  \hline
Total         & 10.65 & 10.49 & 7.06 & 7.82 & 6.97 & 7.42 & 10.08 & 10.90\\ \hline \hline

\end{tabular}
\end{center}
\end{table}

\begin{table}[!htb]
\caption{\label{tab:syste-psi2skstar}
Breakdown of contributions to the systematic errors for the \psitwos \Kstar channels. 
All values are expressed relative to the measured branching fractions, in percent.}
\begin{center}
\begin{tabular}{l|c|c|c|c}
  & \KS \piz  & \Kp \pim & \KS \pim & \Kp \piz \\ \hline
B counting    & 1.1 & 1.1 & 1.1 & 1.1 \\
Tracking      & 2.6 & 5.2 & 3.9 &3.9 \\
\KS           & 2.0 & - & 1.9 & - \\
Neutral       & 7.2 & - & - & 6.2 \\
Second BF     & 11.74 & 11.72 & 11.72 & 11.72 \\
PID           & 0.97 & 1.62 & 0.21 & 0.52 \\
Background    & 9.36 & 7.83 & 8.59 & 10.82 \\
Polarization  & 6.11 & 4.72 & 4.67 & 7.19 \\
Mc Statistics & 3.45 & 1.68 & 2.72 & 2.27 \\  \hline
Total         & 18.42 & 15.95 & 16.13 & 19.14 \\ \hline \hline
\end{tabular}
\end{center}
\end{table}

\begin{table}[!htb]
\caption{\label{tab:syste-chiconekstar}
Breakdown of contributions to the systematic errors for the \chicone \Kstar channels. 
All values are expressed relative to the measured branching fractions, in percent.}
\begin{center}
\begin{tabular}{l|c|c|c|c}
  & \KS \piz  & \Kp \pim & \KS \pim & \Kp \piz \\ \hline
B counting    & 1.1 & 1.1 & 1.1 & 1.1 \\
Tracking      & 2.6 & 5.2 & 3.9 &3.9 \\
\KS           & 1.1 & - & 1.1 & - \\
Neutral       & 9.4 & 2.5 & 2.8 & 8.3 \\
Second BF     & 10.4 & 10.4 & 10.4 & 10.4 \\
PID           & 0.35 & 1.64 & 0.85 & 1.23 \\
Background    & 30.08 & 16.76 & 24.25 & 246.53 \\
Polarization  & 8.27 & 5.86 & 6.81 & 8.61 \\
MC statistics & 2.06 & 1.38 & 1.61 & 1.80 \\  \hline
Total         & 34.40 & 21.50 & 27.78 & 247.08\\ \hline \hline
\end{tabular}
\end{center}
\end{table}

\vskip 0.5cm
The systematic uncertainty on the number of \BB events is 1.1$\%$. It is common 
to all the branching fraction measurements. The secondary branching fractions and 
their errors have been taken from \cite{ref:pdg2004}. 

For the tracking efficiency, 
we have used a flat correction of 0.8$\%$ per track with an associated error of 
1.3$\%$ per track. The \KS efficiency corrections have been determined by the use of 
control samples and its errors from varying the \KS selection. The resulting 
error on the \KS efficiency varies from 0.8$\%$ to 2.0$\%$ depending on the channel. 
The uncertainty on the detection and energy measurement of photons is 2.5$\%$ common 
to all channels plus an additional channel-dependent correction. The uncertainty 
on the \piz reconstruction is 5.0$\%$ for all channels plus an additional 
channel-dependent correction. For the particle identification efficiency correction, 
we have assigned a systematic error equal to half of the correction. 
The overall selection efficiency depends on the angular distribution used in the 
simulation for the decay. It can be written as $\epsilon=a+|A_0|^2b$, where $a$ and $b$ 
are functions of the \Kstar helicity angle, 
$a = 3/4 \int (1-\cos^2\theta_{\Kstar}) \epsilon(\theta_{\Kstar}) 
\sin(\theta_{\Kstar})d\theta_{\Kstar}$ and 
$b = 3/4 \int (3\cos^2\theta_{\Kstar}-1) \epsilon(\theta_{\Kstar}) 
\sin(\theta_{\Kstar})d\theta_{\Kstar}$, 
and $|A_0|$ is the (unknown) 
fraction of the longitudinal \Kstar polarization \cite{ref:a0}. We estimate the uncertainty 
due to our ignorance on the value of $|A_0|$ and derive an associated systematical error 
varying from 3.4 to 8.6$\%$, depending on the channel. 
The systematic error due to the finite size of the Monte Carlo statistics varies from 
1.23 to 3.45$\%$. Interference effects between \Kstar events described by a P-wave 
and non-resonant events described by an S-wave have been considered \cite{ref:silvano}. 
The interference term is proportional to the fraction of non-resonant events with respect 
to the number of signal events \cite{ref:silvano}. However, this fraction is small for 
all channels. 
Furthermore, a large systematic uncertainty (see below) has been assigned to the number 
of non-resonant events. Thus, no additional systematic uncertainty due to interference 
effects has been introduced.

In the default fit for the determination of the combinatorial background, the 
shape parameter of the Argus function is not constrained. To determine a systematic 
error, a second fit with the shape parameter of the Argus function fixed to the 
value obtained from fitting the data in the \DeltaE sideband region was performed. 
We have taken as the systematic uncertainty on the combinatorial background 50$\%$ 
of the difference between the combinatorial background contribution obtained 
from the default fit and from the second fit. For the feed-across 
component to the peaking background we have assigned as the systematic error, the 
uncertainty of the corresponding branching fractions, taken from \cite{ref:pdg2004}. 
For the inclusive charmonium contribution to the peaking background, we have 
assigned a 50$\%$ error, accounting for the poor knowledge of the branching 
fractions of the contributing decay modes.

The systematic uncertainties for all modes are listed in Tables \ref{tab:syste-jpsikstar}, 
\ref{tab:syste-psi2skstar}, \ref{tab:syste-chiconekstar} and \ref{tab:syste-kpks}.

\begin{table}
\caption{\label{tab:syste-kpks}
Breakdown of contributions to the systematic errors for the \KS and \Kp channels. 
All values are expressed relative to the measured branching fractions, in percent.}
\begin{center}
{\small
\begin{tabular}{l|cc|cc|cc|cc|cc|cc} 
  & \multicolumn{2}{c}{\jpsi \KS}  & \multicolumn{2}{c}{\jpsi \Kp} & 
    \multicolumn{2}{c}{\psitwos \KS} & \multicolumn{2}{c}{\psitwos \Kp} & 
    \multicolumn{2}{c}{\chicone \KS} & \multicolumn{2}{c}{\chicone \Kp}     \\

 &\epem&\mumu&\epem&\mumu&\epem&\mumu&\epem&\mumu&\epem&\mumu&\epem&\mumu \\ \hline

B counting & \multicolumn{2}{|c|}{1.1} & \multicolumn{2}{|c|}{1.1} & 
\multicolumn{2}{|c|}{1.1} & \multicolumn{2}{|c}{1.1} & 
 \multicolumn{2}{|c|}{1.1} & \multicolumn{2}{|c}{1.1} \\

Tracking & \multicolumn{2}{|c|}{2.6} & \multicolumn{2}{|c|}{3.9} & 
\multicolumn{2}{|c|}{2.6} & \multicolumn{2}{|c}{3.9}  & 
 \multicolumn{2}{|c|}{2.6} & \multicolumn{2}{|c}{3.9} \\

\KS & \multicolumn{2}{|c|}{0.7} & \multicolumn{2}{|c|}{-} & 
\multicolumn{2}{|c|}{1.0} & \multicolumn{2}{|c}{-}  & 
 \multicolumn{2}{|c|}{0.9} & \multicolumn{2}{|c}{-} \\

PID           &0.20&1.55&1.04&1.55&0.22&2.41&0.74&2.08&0.74&1.94&0.82&2.67 \\
Second BF     &1.70&1.71&1.69&1.70&4.11&10.96&4.11&10.96&10.27&10.26&10.27&10.27 \\
Background    &0.22&0.08&0.17&0.03&0.41&0.27&0.57&0.17&6.72&2.96&1.97&1.84 \\
MC statistics &0.56&0.54&1.07&1.01&1.36&1.31&1.88&1.81&1.14&1.11&1.54&1.50 \\ \hline
Total         &3.46&3.75&4.64&4.77&5.28&11.69&6.14&12.01&12.70&11.30&11.40&11.665 \\ \hline \hline

\end{tabular}
}
\end{center}
\end{table}

\section{Physics results}
\label{sec:Physics}

The branching fractions that have been measured separately for the \jpsi \ra \epem and 
\jpsi \ra \mumu decay modes were found to be in good agreement. 
They have therefore been combined. Furthermore, for \Kstar channels, the branching fractions 
from the two neutral sub-modes \KS \piz and \Kp \pim have been averaged together, and the 
branching fractions from the two charged sub-modes \KS \pip and \Kp \piz 
have been averaged together as well. 
The branching fraction measurements are summarized in Table \ref{tab:results}.

\begin{table}[h]
\caption{\label{tab:results}
Measured branching fractions for exclusive decays of \B mesons to 
charmonium and kaon final states. The first error is statistical and the second 
systematic.}
\begin{center}
\begin{tabular}{lc}  \hline \hline
Channel & Branching fraction ($\times 10^{-4}$) \\ \hline

\Bz \ra \jpsi \Kstarz      &  12.92$\pm$0.25$\pm$0.75 \\ 
\Bp \ra \jpsi \Kstarp      &  14.34$\pm$0.36$\pm$0.94 \\ 
\Bp \ra \jpsi \Kp          &  10.55$\pm$0.15$\pm$0.48 \\ 
\Bz \ra \jpsi \Kz          &  8.73$\pm$0.23$\pm$0.30 \\ 

\Bz \ra \psitwos \Kstarz   &  6.65$\pm$0.57$\pm$1.00 \\ 
\Bp \ra \psitwos \Kstarp   &  6.03$\pm$0.85$\pm$0.91 \\ 
\Bp \ra \psitwos \Kp       &  6.31$\pm$0.33$\pm$0.44 \\ 
\Bz \ra \psitwos \Kz       &  6.60$\pm$0.60$\pm$0.46 \\ 

\Bz \ra \chicone \Kstarz   &  3.19$\pm$0.37$\pm$0.64 \\ 
\Bp \ra \chicone \Kstarp   &  2.89$\pm$0.69$\pm$0.93  \\
\B \ra \chicone \Kp        &  5.72$\pm$0.24$\pm$0.64 \\ 
\Bz \ra \chicone \Kz       &  4.56$\pm$0.39$\pm$0.51 \\ \hline \hline

\end{tabular}
\end{center}
\end{table}

From these measurements, we have determined the ratios of charged to neutral 
branching fractions. We have assumed a value of one for the charged to neutral 
\B meson production rate at the $\Upsilon(4S)$. 
The results are presented in Table \ref{tab:ratios}. 
The systematic uncertainties of the ratios have been determined by taking into account the 
correlations of the errors between the branching fractions.

\begin{table}[h]
\caption{\label{tab:ratios}
Results for ratios of charged to neutral braching fractios. 
The first error is statistical and the second systematic. }
\begin{center}
\begin{tabular}{lc} \hline \hline
Ratio  &  Result \\ \hline
$\BR(\Bp \ra \jpsi \Kp)/\BR(\Bz \ra \jpsi \Kz)$               &  1.21$\pm$0.04$\pm$0.04  \\ 
$\BR(\Bp \ra \psitwos \Kp)/\BR(\Bz \ra \psitwos \Kz)$         &  0.95$\pm$0.10$\pm$0.03  \\ 
$\BR(\Bp \ra \chicone \Kp)/\BR(\Bz \ra \chicone \Kz)$          &  1.25$\pm$0.12$\pm$0.07  \\ 
$\BR(\Bp \ra \jpsi \Kstarp)/\BR(\Bz \ra \jpsi \Kstarz)$       &  1.11$\pm$0.04$\pm$0.08  \\ 
$\BR(\Bp \ra \psitwos \Kstarp)/\BR(\Bz \ra \psitwos \Kstarz)$ &  0.91$\pm$0.15$\pm$0.11  \\ 
$\BR(\Bp \ra \chicone \Kstarp)/\BR(\Bz \ra \chicone \Kstarz)$ &  0.91$\pm$0.24$\pm$0.31  \\ 
\hline \hline
\end{tabular}
\end{center}
\end{table}

\noindent
Combining all the  measurements, we obtain:

\begin{eqnarray}
{ \BR(\Bp \ra \rm charmonium \ \kaon^{(*)+})  \over \BR(\Bz \ra charmonium \ \kaon^{(*)0}) } = 
1.14 \pm 0.02 \pm 0.03
\end{eqnarray}

Assuming isospin invariance in the \B \ra charmonium \kaon (\Kstar) decays we can compute 
our own value for the charged to neutral \B meson production. Using the ratio of the charged 
to neutral \B meson lifetimes 
$\tau_{\Bp} / \tau_{\Bz} = 1.086 \pm 0.017$ \cite{ref:pdg2004}), we obtain:

\begin{eqnarray}
R^{+/0} = { \Gamma( \Upsilon(4S) \ra \BpBm)  \over \Gamma( \Upsilon(4S) \ra \BzBzb ) } = 1.05\pm0.04
\end{eqnarray}

\noindent
We also determine the ratio of branching fractions for a vector to a pseudo-scalar light 
meson: $\BR(\Bz \ra \psi \Kstarz) / \BR(\Bz \ra \psi \Kz)$ and 
$\BR(\Bp \ra \psi \Kstarp) / \BR(\Bp \ra \psi \Kp)$ for the three charmonium states 
$\psi$ = \jpsi, \psitwos and \chicone. The results are presented in Table \ref{tab:vectscal}. 
For each of the charmonium states, we also present the average of the charged and neutral 
measurements.

\begin{table}[h]
\caption{\label{tab:vectscal}
Results for  ratio of the branching fractions for a vector (\Kstar ) versus pseudoscalar 
(\kaon ) light meson. The first error is statistical and the second systematic. }
\begin{center}
\begin{tabular}{lr} \hline \hline
Ratio  &  Result \\ \hline
$\BR(\Bz \ra \jpsi \Kstarz)/\BR(\Bz \ra \jpsi \Kz)$       &  1.48$\pm$0.05$\pm$0.07  \\ 
$\BR(\Bp \ra \jpsi \Kstarp)/\BR(\Bp \ra \jpsi \Kp)$       &  1.36$\pm$0.04$\pm$0.08  \\ 
$\BR(\B \ra \jpsi \Kstar)/\BR(\B \ra \jpsi \kaon)$ &  1.42$\pm$0.03$\pm$0.05  \\ \hline 

$\BR(\Bz \ra \psitwos \Kstarz)/\BR(\Bz \ra \psitwos \Kz)$ &  1.01$\pm$0.13$\pm$0.09  \\ 
$\BR(\Bp \ra \psitwos \Kstarp)/\BR(\Bp \ra \psitwos \Kp)$ &  0.96$\pm$0.14$\pm$0.09  \\ 
$\BR(\B \ra \psitwos \Kstar)/\BR(\B \ra \psitwos \kaon)$ &  
                                                             0.99$\pm$0.10$\pm$0.06  \\ \hline 

$\BR(\Bz \ra \chicone \Kstarz)/\BR(\Bz \ra \chicone \Kz)$ &  0.70$\pm$0.10$\pm$0.12  \\ 
$\BR(\Bp \ra \chicone \Kstarp)/\BR(\Bp \ra \chicone \Kp)$ &  0.51$\pm$0.12$\pm$0.15  \\ 
$\BR(\B \ra \chicone \Kstar)/\BR(\B \ra \chicone \kaon)$ &  
                                                          0.62$\pm$0.08$\pm$0.09  \\ \hline \hline
\end{tabular}
\end{center}
\end{table}

Finally, charge asymmetries have been measured. The branching fractions for 
positively and negatively charged \B mesons have been determined using the method 
described above. The selection efficiencies have been determined separately.

\begin{eqnarray}
{ \BR(\Bp \ra \jpsi \Kp) -  \BR(\Bm \ra \jpsi \Km) \over 
  \BR(\Bp \ra \jpsi \Kp) +  \BR(\Bm \ra \jpsi \Km) } = -0.029 \pm 0.014 \pm 0.010
\end{eqnarray}

\vskip-0.3cm
\begin{eqnarray}
{ \BR(\Bp \ra \jpsi \Kstarp) -  \BR(\Bm \ra \jpsi \Kstarm) \over 
  \BR(\Bp \ra \jpsi \Kstarp) +  \BR(\Bm \ra \jpsi \Kstarm) } = 0.045 \pm 0.025 \pm 0.011
\end{eqnarray}

\vskip-0.3cm
\begin{eqnarray}
{ \BR(\Bp \ra \psitwos \Kp) -  \BR(\Bm \ra \psitwos \Km) \over 
  \BR(\Bp \ra \psitwos \Kp) +  \BR(\Bm \ra \psitwos \Km) } = 0.059 \pm 0.051 \pm 0.021
\end{eqnarray}

\vskip-0.3cm
\begin{eqnarray}
{ \BR(\Bp \ra \psitwos \Kstarp) -  \BR(\Bm \ra \psitwos \Kstarm) \over 
  \BR(\Bp \ra \psitwos \Kstarp) +  \BR(\Bm \ra \psitwos \Kstarm) } = -0.063 \pm 0.137 \pm 0.050
\end{eqnarray}

\vskip-0.3cm
\begin{eqnarray}
{ \BR(\Bp \ra \chicone \Kp) -  \BR(\Bm \ra \chicone \Km) \over 
  \BR(\Bp \ra \chicone \Kp) +  \BR(\Bm \ra \chicone \Km) } = 0.011 \pm 0.042 \pm 0.017
\end{eqnarray}

\vskip-0.3cm
\begin{eqnarray}
{ \BR(\Bp \ra \chicone \Kstarp) -  \BR(\Bm \ra \chicone \Kstarm) \over 
  \BR(\Bp \ra \chicone \Kstarp) +  \BR(\Bm \ra \chicone \Kstarm) } = -0.403 \pm 0.309 \pm 0.237
\end{eqnarray}

\section{Summary}
\label{sec:Summary}
We have presented preliminary results of branching fraction measurements of exclusive 
\B decays to charmonium and \kaon or \Kstar. The charmonium mesons considered were 
\jpsi, \psitwos and \chicone. Our results for \jpsi and \psitwos are in good agreement 
with previous measurements \cite{ref:pdg2004} with comparable or superior precision. 
Our \chicone results have much better precision and the \Bp \ra \chicone \Kstarp 
branching fraction was measured for the first time. Assuming isopin invariance, we find 
the ratio of charged to neutral \B meson production on the $\Upsilon$(4S) 
resonance to be compatible with unity within two standard deviations. No 
direct \CP violation has been observed from the measurements of charge asymmetries as 
we found them to be compatible with zero.

\section{Acknowledgments}
\label{sec:Acknowledgments}

We are grateful for the 
extraordinary contributions of our \pep2\ colleagues in
achieving the excellent luminosity and machine conditions
that have made this work possible.
The success of this project also relies critically on the 
expertise and dedication of the computing organizations that 
support \babar.
The collaborating institutions wish to thank 
SLAC for its support and the kind hospitality extended to them. 
This work is supported by the
US Department of Energy
and National Science Foundation, the
Natural Sciences and Engineering Research Council (Canada),
Institute of High Energy Physics (China), the
Commissariat \`a l'Energie Atomique and
Institut National de Physique Nucl\'eaire et de Physique des Particules
(France), the
Bundesministerium f\"ur Bildung und Forschung and
Deutsche Forschungsgemeinschaft
(Germany), the
Istituto Nazionale di Fisica Nucleare (Italy),
the Foundation for Fundamental Research on Matter (The Netherlands),
the Research Council of Norway, the
Ministry of Science and Technology of the Russian Federation, and the
Particle Physics and Astronomy Research Council (United Kingdom). 
Individuals have received support from 
CONACyT (Mexico),
the A. P. Sloan Foundation, 
the Research Corporation,
and the Alexander von Humboldt Foundation.

\end{document}